\documentclass[12pt]{article}
\usepackage{epsfig}
\newcommand{\nn}{\nonumber}

\newcommand{\be}{\begin{equation}}
\newcommand{\ee}{\end{equation}}
\newcommand{\bea}{\begin{eqnarray}}
\newcommand{\eea}{\end{eqnarray}}

\newcommand{\aprle}{\stackrel{<}{{}_\sim}}

\newcommand{\evb}{ {\rm eV$^2$} }

\begin{document}
%
\thispagestyle{empty}
\begin{flushright}
{\tt hep-ph/0107231}\\
{ROMA1-TH/2001-1320}
\end{flushright}
\vspace*{1cm}
\begin{center}
{\Large{\bf Telling three from four neutrinos at the Neutrino Factory} }\\
\vspace{.5cm}
A. Donini$^{\rm a,}$\footnote{andrea.donini@roma1.infn.it},
M. Lusignoli$^{\rm b,}$\footnote{maurizio.lusignoli@roma1.infn.it}
and D. Meloni$^{\rm b,}$\footnote{davide.meloni@roma1.infn.it}
 
\vspace*{1cm}
$^{\rm a}$ I.N.F.N., Sezione di Roma I and Dip. Fisica, 
Universit\`a di Roma ``La Sapienza'', 
P.le A. Moro 2, I-00185, Rome, Italy \\ 
$^{\rm b}$ Dip. Fisica, Universit\`a di Roma ``La Sapienza''
and I.N.F.N., Sezione di Roma I, P.le A. Moro 2, I-00185, Rome, Italy

\end{center}
\vspace{.3cm}
\begin{abstract}
\noindent
We upgrade the study of the physical reach of a Neutrino Factory
considering the possibility to distinguish a three (active) neutrino 
oscillation scenario from the scenario in which a light sterile 
neutrino is also present. The distinction is easily performed in the 
so--called 2+2 scheme, but also in the more problematic 3+1 scheme it 
can be attained in some regions of the parameter space. We also discuss 
the CP violating phase determination, showing that the effects of a 
large phase in the three--neutrino theory cannot be reproduced in a 
four--neutrino, CP conserving, model. 
\end{abstract}

\newpage

\section{Introduction}
\label{sec:intro}

Indications in favour of neutrino oscillations 
\cite{Pontecorvo:1957yb,Maki:1962mu,Pontecorvo:1968fh,Gribov:1969kq}
have been 
obtained both in solar neutrino 
\cite{Cleveland:1998nv,Fukuda:1996sz,Hampel:1999xg,Abdurashitov:1999zd,Suzuki:1999cy,Ahmad:2001an}
and atmospheric neutrino 
\cite{Fukuda:1994mc,Becker-Szendy:1995vr,Fukuda:1999ah,Allison:1999ms,Ambrosio:1998wu} 
experiments.
The atmospheric neutrino data require 
$\Delta m_{atm}^2 \sim (1.6 - 4)\;10^{-3}$ \evb \cite{Toshito:2001dk}
whereas the solar neutrino data prefer $\Delta m_\odot^2 \sim 10^{-10}$
or $10^{-7} - 10^{-4}\;{\rm eV}^{2}$, depending on the particular oscillation 
solution chosen for the solar neutrino deficit. 
The LSND data \cite{Athanassopoulos:1998pv,Aguilar:2001ty}, 
on the other hand, would indicate 
a $\nu_\mu \to \nu_e$ oscillation with a third, very distinct, 
neutrino mass difference: $\Delta m_{LSND}^2 \sim 0.3 - 6\;{\rm eV}^2$. 
The LSND evidence in favour of neutrino oscillation has not been confirmed 
by other experiments so far \cite{Kleinfeller:2000em}; 
the MiniBooNE experiment \cite{Church:1997jc}
will be able to do it in the near future.
If MiniBooNE will confirm the LSND results, we would face
three independent evidence for neutrino oscillations characterized
by squared mass differences quite well separated. 
To explain the whole ensemble of data four different light neutrino species 
are needed; the new light neutrino is denoted as sterile, 
since it must be an electroweak singlet
to comply with the strong bound \cite{Caso:2000tc} on the $Z^0$ invisible 
decay width. 
We stress that all the present experimental results (LSND included) 
cannot be explained with three massive light neutrinos only,
as it has been shown with detailed calculations in \cite{Fogli:1999zq}.

In order to improve our present knowledge about neutrino 
masses and mixings, it has been proposed to build a Neutrino Factory 
\cite{Geer:1998iz,DeRujula:1999hd}, with two long baseline experiments 
running at the same time at two different distances. One of its main 
goal would be the discovery of leptonic CP violation and, possibly, 
its study \cite{Barger:2000fs,Bueno:2000wb,Dick:1999ed,Cervera:2000kp}.
Previous analyses \cite{Albright:2000xi,Blondel:2000gj} 
on the foreseeable outcome of experiments at a 
Neutrino Factory have shown that the determination of the 
two still unknown parameters in the three--neutrino mixing matrix, 
$\theta_{13}$ and $\delta$, will be possible, while with 
conventional neutrino beams it would not be so \cite{Barger:2001qd}.
The scenario in which the LSND result has been confirmed  
(and therefore a short baseline experiment to study the four--neutrino 
parameter space is required) has been considered in 
\cite{Donini:1999jc,Donini:2001xy}.

If the result of the MiniBooNE experiment will not be conclusive, 
it seems anyhow quite relevant to understand if three-- and four--family 
models are distinguishable, and to what extent. Moreover, it is a sensible 
question to ask if the effect of CP violation in a three--family world could be 
mimicked in a four--family, CP conserving, neutrino scenario.

There are two very different classes of spectra with four massive neutrinos: 
three almost degenerate neutrinos and an isolated fourth one, 
or two pairs of almost degenerate neutrinos divided by the large
LSND mass gap. The two classes of mass spectra are called for obvious reasons 
the 3+1 and 2+2 scheme, respectively. The experimental results were strongly
in favour of the 2+2 scheme \cite{Bilenkii:1999ny} until the latest
LSND results \cite{Aguilar:2001ty}. 
The new analysis of the experimental data results in a shift of the
allowed region towards smaller values of the mixing angle, 
$\sin^2 (2 \theta_{LSND})$, reconciling the 3+1 scheme with exclusion 
bounds coming from other reactor and accelerator experiments 
\cite{Dydak:1984zq,Stockdale:1985ce,Declais:1995su,Eskut:2001de,Astier:2001yj}.
As a consequence, the 3+1 model is (marginally) compatible with the 
data \cite{Barger:2000ch,Yasuda:2000xs,Giunti:2001ur,Peres:2001ic}. 
Although the 2+2 scheme gave a better fit to those data (as was shown
in \cite{Grimus:2001mn}), the recent SNO results \cite{Ahmad:2001an} 
will certainly restrict the allowed parameter region and give a 
considerably worse fit for this model. This is understandable from the analysis 
presented in refs.\cite{Gonzalez-Garcia:2001hs,Gonzalez-Garcia:2001uy}, 
prior to the SNO data publication, since the SMA solution for solar neutrinos, 
which was more easily compatible with the limit on sterile components 
in atmospheric oscillations, is at present strongly disfavoured 
(see also \cite{Barger:2001zs,Bahcall:2001zu}).

It will certainly be impossible to tell a four--family 3+1 
from a three--family scheme in the case of very small active--sterile 
mixing angles: the 3+1 model reduces to a 
three--family scheme for vanishing mixing with the isolated state. 
A discrimination will 
therefore be possible only if those angles are large enough. 
On the other hand, the four--family 2+2 scheme has the LSND result 
built--in (in  the sense that it cannot stand if LSND is not 
confirmed by MiniBooNE) and it does not go smoothly into a 
three--family model. 

We address the problem of distinction between four and 
three families in this paper. In order to do that, we generated 
``experimental data'' \cite{Cervera:2000kp} according to a four--family 
theoretical scheme without CP violation and tried to fit these data 
with a three--family model having a free CP violating phase $\delta$. 
As a cross--check, we also proceeded in the opposite direction, fitting with  
four--family formulae the data generated in a three--family theory. 
The data have been generated at three distances, $L$ = 732, 3500, 
7332~Km, using the detector and machine parameters as defined in 
\cite{Cervera:2000kp}. 

The main conclusions of this analysis are: \\
\begin{itemize}
    \item 
    The data generated in four--family without CP violation can 
    be described by the three--family formulae in some particular 
    zones of the four--family parameter space, not restricted 
    to the obvious case of very small angles. These zones are reduced 
    in size for increasing gap--crossing angles.
    \item 
    Combining data taken at different distances the zones in 
    which a description with the other theory is successful are 
    generally reduced.
    \item 
    An increase of the energy resolution of the detector may have a 
    positive effect (reducing the zones of ambiguity), but only in the 
    case of rather large gap--crossing angles.
    \item Whenever a particular zone in the four--family parameter
    space gives acceptable $\chi^{2}$ when fitted in the three-family 
    scheme, the fitted value of the three--family CP 
    violating phase, $\delta$, is generally not large. In particular, 
    this is true for $L = 3500$ Km, 
    whereas for $L = 7332$ Km the determination of $\delta$ is somewhat looser, 
    due to the overwhelming matter effects. For the shortest baseline, 
    $L = 732$ Km, we have the largest spread in the values of $\delta$, 
    although the most probable value is still close to zero. 
    \item 
    The cross--check, fitting three--family generated data in a 
    four--family model, is consistent with the results previously 
    obtained. Fitting data generated with a CP phase close to 90$^\circ$ 
    in a CP conserving 3+1 theory is practically impossible if one combines 
    data at two different distances.  
    \item In the 2+2 scheme (as opposed to 3+1) the ambiguity with a 
    three--neutrino theory is essentially absent. 
\end{itemize}

In Section \ref{sec:form} we present the relevant formulae for the oscillation 
probabilities; in Section \ref{sec:3or4} we discuss the possibility to have 
a fit in the three--neutrino theory of ``data'' generated with the 3+1 
model; in Section \ref{sec:delta} we give a discussion of the previous fits, 
particularly regarding the value of the CP violating phase  
$\delta$. Section \ref{sec:concl} is devoted to the conclusions.

\section{Three and Four Neutrino Oscillations}
\label{sec:form}

Our ordering of the mass eigenstates corresponds to increasing 
absolute value of the mass, $m_i^2 < m_j^2$ for $i<j$. 
We restrict ourselves, for simplicity, to the hierarchical 3+1 scheme 
($\Delta m^2_{21} = \Delta m^2_\odot$, $\Delta m^2_{32} = \Delta m^2_{atm}$
and $\Delta m^2_{43} = \Delta m^2_{LSND}$) 
and to the so-called ``class II-B'' 2+2 scheme 
($\Delta m^2_{21} = \Delta m^2_\odot$, $\Delta m^2_{32} = \Delta m^2_{LSND}$
and $\Delta m^2_{43} = \Delta m^2_{atm}$). 
All the other spectra may be obtained changing the sign to one or more
of the squared mass differences.

The Pontecorvo-Maki-Nakagawa-Sakata (PMNS) mixing matrix 
\cite{Pontecorvo:1957yb,Maki:1962mu,Pontecorvo:1968fh,
Gribov:1969kq} for three (active) neutrinos is usually 
parameterized in terms of three angles $\theta_{ij}$ and one CP violating 
phase $\delta$ (for Majorana neutrinos, two additional CP violating 
phases appear, but with no effect on the oscillations):
\be
U_{PMNS}^{(3)}=U_{23}(\theta_{23})\;U_{13}(\theta_{13}\,,\,\delta)\;
U_{12}(\theta_{12})\;.
\label{MNS3}
\ee

Four neutrino oscillations imply a $4 \times 4$
mixing matrix, with six rotation angles and three phases
(for Majorana neutrinos, three additional phases are allowed). 
It is convenient to use two slightly different parameterizations.
In the 3+1 case we write the mixing matrix as
\be
U_{PMNS}^{(3+1)} = U_{14} (\theta_{14})\;U_{24} (\theta_{24})\;U_{34} (\theta_{34})\; 
   U_{23} (\theta_{23}\,,\,\delta_3)\;  U_{13} 
   (\theta_{13}\,,\,\delta_2)\; 
    U_{12} (\theta_{12}\,,\,\delta_1)\; ,
\label{ourpar31}
\ee
which reduces to the three--family case in the limit 
$\theta_{i4} \to 0$.

A more convenient parameterization for the 2+2 case, as shown in 
\cite{Donini:1999jc}, is
\be
U_{PMNS}^{(2+2)} = U_{14} (\theta_{14})\;U_{13} (\theta_{13})\;U_{24} (\theta_{24})\; 
   U_{23} (\theta_{23}\,,\,\delta_3) \; U_{34} 
   (\theta_{34}\,,\,\delta_2)\; 
    U_{12} (\theta_{12}\,,\,\delta_1)\; .
\label{ourpar22}
\ee

The general formula for the oscillation probability is well known:
\bea
P(\nu_{\alpha} \to \nu_{\beta}) &=& \delta_{\alpha \beta} - 4\,\sum_{i>j} 
{\rm Re} [U_{\alpha,i} U_{\beta,j}
U_{\alpha,j}^{*} U_{\beta,i}^{*}] \;\sin^{2}\left({\Delta m^{2}_{ij}\;L  
\over 4\,E} \right) \nn \\
&\mp& 2\,\sum_{i>j} {\rm Im} [U_{\alpha,i} U_{\beta,j} U_{\alpha,j}^{*} U_{\beta,i}^{*}] 
\; \sin \left({\Delta m^{2}_{ij}\;L \over 2\,E} \right)
\label{prob}
\eea
(the lower sign applies to the antineutrino case).

\begin{figure}[h!]
\begin{center}
\epsfxsize6.5cm\epsffile{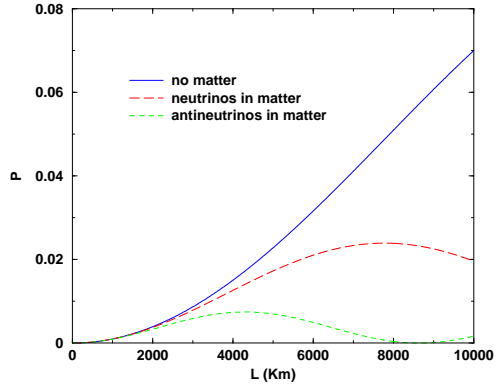} 
\caption{\label{matagain} Transition probability in vacuum and in 
matter (for constant $\rho$= 2.8 g cm$^{-3}$) in the $\nu_e \to \nu_\mu$ 
($\bar\nu_e \to \bar\nu_\mu$)  
channel for $E = 38 \, GeV$. The following parameters have been 
assumed: $\theta_{12}=22.5^\circ$, $\theta_{23}=45^\circ$, 
$\theta_{13}=13^\circ$, and vanishing CP violating phase.}
\label{fig:matter}
\end{center}
\end{figure}

In all of our numerical calculations we used the exact formulae, 
including matter effects, that are only important 
for pathlengths greater than about 2000 Km (see 
fig.~\ref{fig:matter} for an example in the three--family model).

First, we recall the three--family oscillation probabilities 
in vacuum at distances short enough, so that the oscillations of 
longest length, due to $\Delta m^2_\odot$, do not have time to 
develop \cite{DeRujula:1980yy}. 
In the three--neutrino case, neglect of solar $\Delta m^{2}$ entails no CP 
violation, and the oscillation probabilities are:  
\bea
P_3 (\nu_{e} \to \nu_{\mu}) &=& \sin^{2}(\theta_{23}) \; 
\sin^{2}(2\theta_{13}) \;
 \sin^{2}\left({\Delta m^{2}_{atm}\;L \over 4\,E}\right)\;, 
 \label{eq:threefam-e-mu} \\
P_3 (\nu_{e} \to \nu_{\tau}) &=& \cos^{2}(\theta_{23}) \; 
\sin^{2}(2\theta_{13}) \;
 \sin^{2}\left({\Delta m^{2}_{atm}\;L \over 4\,E}\right)\;, 
 \label{eq:threefam-e-t} \\
P_3 (\nu_{\mu} \to \nu_{\tau}) &=& \cos^{4}(\theta_{13}) \; \sin^{2}(2\theta_{23})
\; \sin^{2}\left({\Delta m^{2}_{atm}\;L \over 4\,E}\right)\;,  
\label{eq:threefam-mu-t} \\ 
P_3 (\nu_{e} \to \nu_{e}) &=& 1 - \sin^{2}(2\theta_{13}) \;
\sin^{2}\left({\Delta m^{2}_{atm}\;L \over 4\,E}\right)\;,  
\label{eq:threefam-e-e} 
\eea
\be
P_3 (\nu_{\mu} \to \nu_{\mu}) = 1 - \left[ \cos^{4}(\theta_{13})\,  
\sin^{2}(2\theta_{23}) + \sin^{2}(\theta_{23})\,  
\sin^{2}(2\theta_{13}) \right]\;
 \sin^{2}\left({\Delta m^{2}_{atm}\;L \over 4\,E}\right)\,. 
 \label{eq:threefam-mu-mu}
\ee

The explicit expressions for the exact formulae in the four-family case 
are quite heavy. It may be of interest to present approximate formulae, 
that allow better intuition at least 
in their range of validity. We consider therefore 
the shortest distance from the neutrino source, $L = 732$ Km 
(the CERN--LNGS and Fermilab--MINOS distance), where matter effects 
are negligible and we present approximate formulae
for the 3+1 four-family mixing in vacuum (see also \cite{Donini:2001xy}).
At this distance the effects of the large  $\Delta m^{2}_{LSND}$ oscillations 
are averaged and the effects of the small $\Delta m^{2}_\odot$ are negligible. 
We also assume, in agreement with present bounds \cite{Barger:2000ch}, 
$\theta_{14}=\theta_{24}=\epsilon$ and we expand in power series in $\epsilon$
to deduce simple expressions to be compared with eqs. 
(\ref{eq:threefam-e-mu}-\ref{eq:threefam-mu-mu}).
The result for vanishing CP violating phases is: 
\bea
P_{3+1} (\nu_e \to \nu_\mu) &=&  4 c_{13}^2 s_{13}^2 s_{23}^2 \sin^2 
\left (\frac {\Delta m^2_{32} L}{4 E} \right )
- \nn \\& & 
      8 \epsilon c_{13}^2 c_{23} s_{13} s_{23} (s_{13} + c_{13} s_{23}) s_{34}
      \sin^2 \left (\frac {\Delta m^2_{32} L}{4 E} \right )+O(\epsilon^2)\;, 
      \label{eq:fourfam-e-mu} \\
P_{3+1} (\nu_e \to \nu_\tau) &=& 4 c_{34}^2 c^{2}_{23} \; 
s^{2}_{13} c^{2}_{13} 
 \sin^{2}\left({\Delta m^{2}_{32}\;L \over 4\,E}\right)
+ \nn \\& & 
      4 \epsilon c_{13} c_{23} s_{13} (1 -2 c_{13}^2 c_{23}^2) c_{34}^2 s_{34}
      \sin^2 \left (\frac {\Delta m^2_{32} L}{4 E} \right )+O(\epsilon^2)\;, 
 \label{eq:fourfam-e-t} \\
P_{3+1} (\nu_\mu \to \nu_\tau)&=&4 c_{13}^4 c_{23}^2 c_{34}^2  s_{23}^2 
\sin^2 \left (\frac {\Delta m^2_{32} L}{4 E} \right )
+ \nn \\& &
      4 \epsilon\,c_{13}^2 c_{34}^2 s_{34} c_{23} s_{23} 
      \left(1-2 c_{13}^2 c_{23}^{2}\right)
      \sin^2 \left (\frac {\Delta m^2_{32} L}{4 E} \right )
      +O(\epsilon^2) \;,
\label{eq:fourfam-mu-t} \\
P_{3+1} (\nu_e \to \nu_e)&=& 1 - 4 c_{13}^2 s_{13}^2 \sin^2 
           \left ( \frac {\Delta m^2_{32} L} {4 E}\right ) + \nn \\
& & 8 \epsilon c_{13} c_{23} s_{13}(c_{13}^2 - s_{13}^2) s_{34}
      \sin^2 \left (\frac {\Delta m^2_{32} L}{4 E} \right )+O(\epsilon^2) \;,
      \label{eq:fourfam-e-e} \\
P_{3+1} (\nu_\mu \to \nu_\mu)&=& 1 - 4 c_{13}^2 s_{23}^2 (1-c_{13}^2 s_{23}^2) 
\sin^2 \left (\frac {\Delta m^2_{32} L}{4 E} \right ) +
\label{eq:fourfam-mu-mu} \\
& & 8 \epsilon c_{13}^2 c_{23} s_{23}[c_{23}^2 +(-1+2 
s_{13}^2)s_{23}^2] s_{34}
      \sin^2 \left (\frac {\Delta m^2_{32} L}{4 E} \right 
      )+O(\epsilon^2) \,. 
\nn 
\eea      
The three-family limit is plainly reached for $\epsilon$ and 
$\theta_{34} \to 0$. 
We notice that the $\theta_{34}$ dependence is generally suppressed with powers of 
$\epsilon$, except in the $\nu_\mu \to \nu_\tau$ channel, that is therefore the most 
sensitive to this angle. From the above formulae it appears that for small enough $\epsilon$ 
it will be extremely difficult to distinguish a four-family 3+1 scheme 
from three families looking at $\nu_\mu \to \nu_e$ transitions, regardless of the value 
of $\theta_{34}$. 

When the ``data'' generated in the 3+1 model are 
fitted in the three--neutrino theory, the result (if 
successful) fixes the three-neutrino parameters to some values, 
that may be called ``effective''. It will be useful for our discussion 
to present explicit formulae for some of them, 
keeping the small $\epsilon$ approximation. They can be easily derived from 
the equations given above, and are:
\bea
\sin^2(2\,\theta_{23}^{\,eff})&=&\sin\,(2\,\theta_{23}) 
\left [ 
\sin\,(2\,\theta_{23}) -4\,\epsilon\,s_{34} \cos\,(2\,\theta_{23}) 
+ O(\epsilon^2) 
\right ]\;,
\label{s-23}\\
\sin^2(2\,\theta_{13}^{\,eff})&=&\sin\,(2\,\theta_{13}) 
\left [ 
\sin\,(2\,\theta_{13}) -4\,\epsilon\,s_{34} c_{23} c_{13}^2 + O(\epsilon^2) 
\right ]\,.
\label{s-13}
\eea

Contrary to the 3+1 model, the other four--neutrino option, 2+2, does 
not have the three--neutrino theory as a limit for small gap--crossing mixing angles. 
In this scheme, some of the oscillation probabilities have an 
expression sufficiently simple to be written down explicitly. In the 
following formulae, the only approximation made is to neglect 
the solar neutrino mass difference $\Delta m^2_\odot$ and 
to average the rapid oscillations due to $\Delta m^2_{LSND}$. 
\bea
P_{2+2} (\nu_e \to \nu_\mu) &=& 
2 \; c_{23}^2 s_{23}^2 c_{13}^2 c_{24}^2 - \nn \\
& & c_{13}^2 c_{23}^2 \left[ \sin^2(2\theta_{34}) 
\left( c_{24}^2 s_{23}^2 - s_{24}^2 \right) + 
\sin(4\theta_{34}) \, c_{24}s_{24}s_{23} \right]  
\sin^2 \left (\frac {\Delta m^2_{43} L}{4 E} \right)\!, \nn \\
\nn \\
P_{2+2} (\nu_e \to \nu_e) &=& 1 - 2 \; c_{23}^2 c_{24}^2 \left(c_{24}^2 
s_{23}^2 + s_{24}^2 \right) - 
\label{four-fam-22} \\
& &  \left[ \sin(2\theta_{34}) 
\left(c_{24}^2 s_{23}^2 - s_{24}^2 \right) + 
2 \cos(2\theta_{34})\,c_{24}s_{24}s_{23} \right]^2 
\sin^2 \left (\frac {\Delta m^2_{43} L}{4 E} \right)\;, \nn \\
\nn \\
P_{2+2} (\nu_\mu \to \nu_\mu)&=& 1 - 2\;c_{13}^2 c_{23}^2 \;(s_{13}^2 + 
c_{13}^2 s_{23}^2) - 4\;c_{13}^4 c_{23}^4 c_{34}^2 s_{34}^2\;
\sin^2 \left (\frac {\Delta m^2_{43} L}{4 E} \right)\,. \nn
\eea
The oscillation probabilities involving tau neutrinos may be easily 
obtained: their expressions are however rather cumbersome, and we omit 
them here for simplicity.

It may be noted that in the limit of vanishing gap--crossing angles 
eq.(\ref{four-fam-22}) reduces to two independent two--neutrinos 
oscillations, $\nu_\mu \leftrightarrow \nu_\tau$ and $\nu_e 
\leftrightarrow \nu_s$ (that is ineffective since $\Delta m^2_\odot = 
0$). Therefore, the possibility of confusion with a three--neutrino 
scheme is practically inexistent, as we will discuss later.

\section{Three or four families?}
\label{sec:3or4}

In a few years from now the LSND results will be confirmed by MiniBooNE 
or not. If the former is the case, a short baseline experiment
that could explore the LSND gap--crossing parameter space would be 
mandatory. In case of a non-conclusive result, 
the three-family mixing model will be considered the most plausible
extension of the Standard Model. If this is the case, long baseline experiments
will be preferred with respect to the (four-family inspired) short baseline ones. 
This is the scenario that we would like to explore, trying to answer 
the following question: will a Neutrino Factory and corresponding detectors
(designed to explore the three-family mixing model) be able to tell three 
neutrinos from four neutrinos?  

We consider the following ``reference set-up'': neutrino beams resulting from 
the decay of $ 2 \times 10^{20} \mu^+$'s and $\mu^-$'s per year in a straight section 
of an $E_\mu = 50$ GeV muon accumulator. An experiment with a realistic 40 Kton 
detector of magnetized iron and five years of data taking 
for each polarity is envisaged. Detailed estimates of the corresponding expected 
backgrounds and efficiencies have been included in the analysis \cite{Cervera:2000vy}. 
Three reference baselines are discussed: 732 Km, 3500 Km and 7332 Km. 
We follow the analysis in energy bins as made in 
\cite{Cervera:2000kp,Donini:2000ky,Burguet-Castell:2001ez}. 
The $\nu_e \to \nu_\mu$ channel, the so--called {\sl golden channel}, 
will be the main subject of our investigations.

Let $N_{i,p}^\lambda$ be the total number of wrong-sign muons detected 
when the factory is run in polarity  $p=\mu^+,\mu^-$, grouped in energy bins 
specified by the index $i$, and three possible 
distances, $\lambda =$  1, 2, 3 (corresponding to $L = 732$ Km, $L = 3500$ Km and 
$L = 7332$ Km, respectively).
In order to simulate a typical experimental situation we generate 
a set of ``data'' $n_{i,p}^\lambda$ as follows: for a given value of the oscillation 
parameters, 
the expected number of events, $N_{i,p}^\lambda$, is computed; 
taking into account backgrounds and detection efficiencies per bin, $b_{i,p}^\lambda$ 
and $\epsilon_{i,p}^\lambda$, we then perform a gaussian 
(or poissonian, depending on the number of events) smearing to mimic the statistical 
uncertainty: 
\begin{eqnarray} 
n_{i,p}^\lambda = \frac{ {\rm Smear} (N_{i,p}^\lambda \epsilon_{i,p}^\lambda + 
b_{i,p}^\lambda) - b_{i,p}^\lambda}{\epsilon_{i,p}^\lambda} \,. 
\end{eqnarray}
Finally, the "data" are fitted to the theoretical expectation as a function of
the neutrino parameters under study, using a $\chi^2$ minimization:
\be
\chi_\lambda^2 = \sum_p \sum_i 
\left(\frac{ n^\lambda_{i,p} \, - \, 
N^{\lambda}_{i,p}}{\delta n^\lambda_{i,p}}\right)^2 \, ,
\label{chi2}
\ee
where $\delta n^\lambda_{i,p}$ is the statistical error for 
$n^\lambda_{i,p}$ (errors on background and efficiencies are neglected).
We verified that the fitting of theoretical numbers to the 
smeared (``experimental'') ones is able to reproduce the values of 
the input parameters.

We then proceed to fit with a three--neutrino model the
``data'' obtained smearing the (3+1)--neutrino theoretical input. We fixed all 
but two of the input parameters (in particular the CP violating 
phases have been put to zero) and calculated the $\chi^{2}$ fitting 
with the 3$\nu$ theory for a grid of values of the two varying parameters. 
In detail, we fixed the values of $\Delta m^{2}_{LSND}=1\;{\rm eV}^2$, 
$\Delta m^{2}_{atm}=2.8\;10^{-3}\;{\rm eV}^2$, $\Delta m^{2}_{\odot}=1\;10^{-4}\;
{\rm eV}^2$,  $\theta_{12}$ = $22.5^{\circ}$ 
and $\theta_{23}$ = $45^{\circ}$, we also assumed  
$\theta_{14} = \theta_{24} = 2^{\circ}, 5^{\circ}, 10^{\circ}$, and 
allowed a variation of 
$\theta_{13}$ from one to ten degrees in steps of $0.5^\circ$ and of $\theta_{34}$ up to 
$50^{\circ}$ \cite{Fogli:2001ir} in steps of $1^\circ$. 
The matter effects have been evaluated assuming an average constant\footnote{
In principle, at distances above 4000-5000 Km a more accurate earth density
profile (such as the PREM \cite{PREM}) should be used instead of the 
constant density approximation adopted here. This point has been
thoroughly discussed in \cite{Ota:2001hf}. The qualitative features
of our results, however, do not depend on this approximation.} 
density $\rho = 2.8\,(3.8)\;{\rm 
g\,cm^{-3}}$ for $L=$732 or 3500 (7332) Km
and a neutron fraction $Y_n=1/2$ along the neutrino path.
In order to obtain results in a form 
that helps intuition, after checking that the conclusions of a 
general fit are similar,
we kept two of the four parameters in the 
three--neutrino model to fixed values\footnote{
Note that these parameter values are equal to the input (3+1 model) values.
This is understandable, since at the considered distances the effects 
of $\Delta m^{2}_{\odot} \neq 0$ are subleading, and the actual value 
of $\theta_{12}$ is therefore almost irrelevant; on the other hand, 
from eq.(\ref{s-23}) it appears that 
the maximal mixing condition ($\theta_{23} = 45^{\circ}$) holds in 
both theories up to corrections that are of second order in the small 
quantities $\theta_{14}$, $\theta_{24}$ and $\theta_{13}$.}, 
$\theta_{12} = 22.5^{\circ}$ 
and $\theta_{23} = 45^{\circ}$ .
We allowed the remaining parameters to vary in the intervals $1^\circ \leq 
\theta_{13} \leq 10^{\circ}$ and -180$^\circ \leq \delta < 180^\circ$. 

\begin{figure}[h!]
\begin{center}
\epsfxsize6cm\epsffile{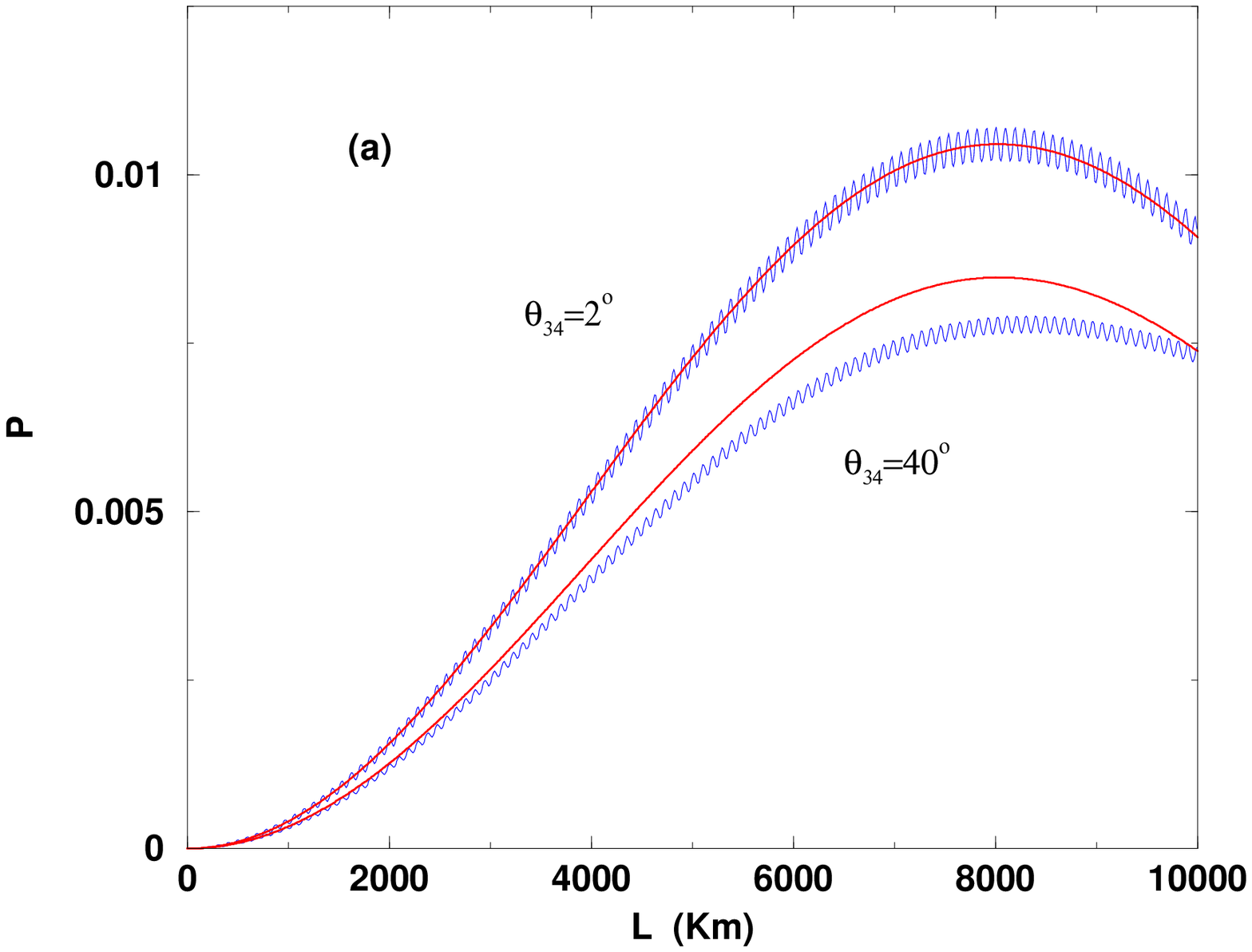} \hspace{0.7truecm} 
\epsfxsize6cm\epsffile{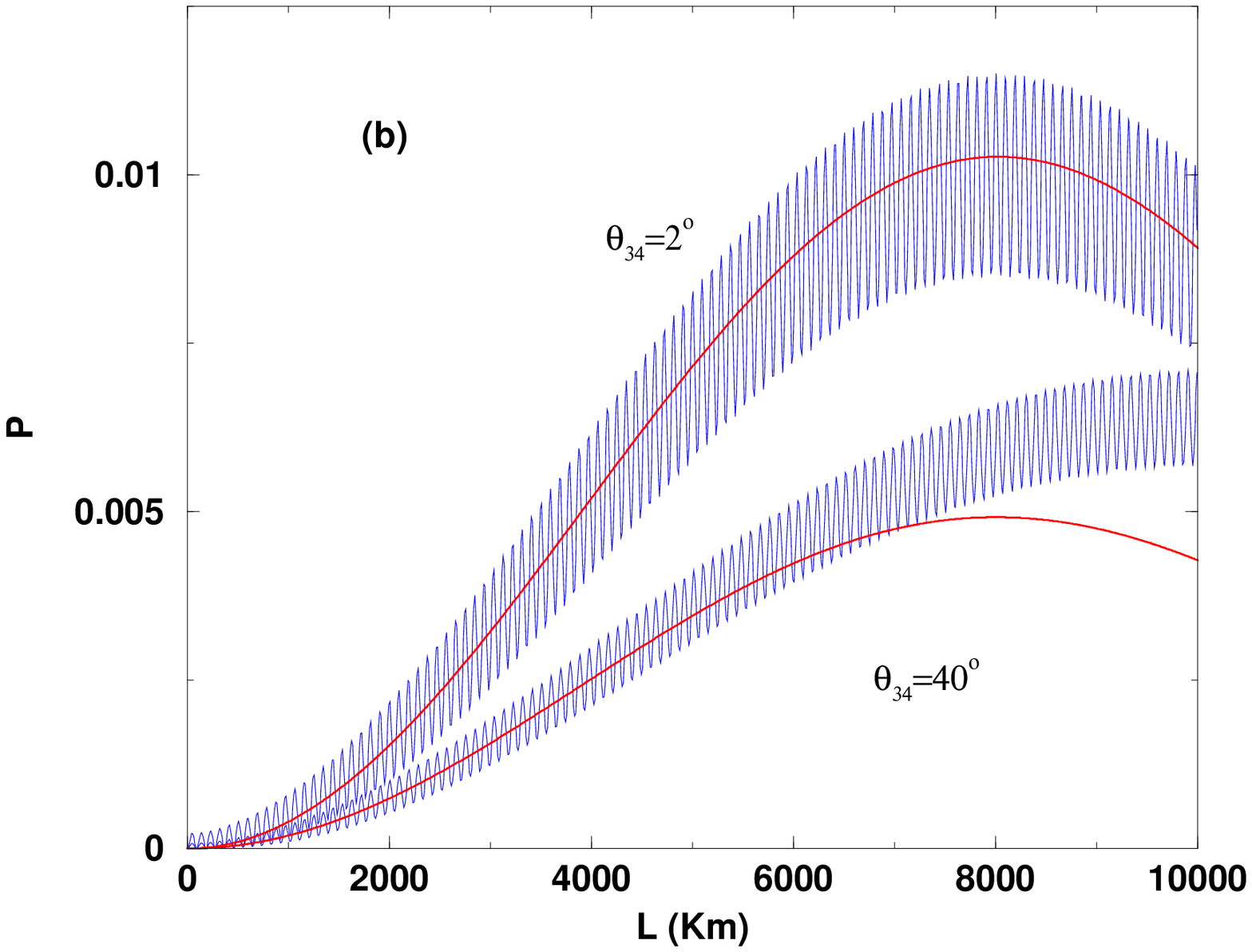}
\caption{ Oscillation probability for $\nu_{e} \to \nu_{\mu}$ as a 
function of the pathlength for $E$ = 38 GeV and (a) $\epsilon 
=2^{\circ}$ or (b) $\epsilon =5^{\circ}$. See text for further details.
\label{fig:prob1}}
\end{center}
\end{figure} 

In order to better understand the outcome of our fits, we consider the 
$\nu_\mu \to \nu_{e}$ conversion probability as a function of the 
pathlength for a fixed neutrino energy $E$ = 38 GeV,
corresponding to the effective mean energy (including the effect of 
the cross section) in a Neutrino Factory with 50 GeV muons, in both scenarios.  
We stick to the simplified approach of 
eqs.(\ref{eq:fourfam-e-mu}-\ref{eq:fourfam-mu-mu}) and assume equal 
values for the angles $\theta_{14}$ = $\theta_{24}$ = $\epsilon$,
after checking that possible sign changes in these quantities do not modify 
the general behaviour of our results at the distances considered.
In fig.~\ref{fig:prob1}a (\ref{fig:prob1}b) this common value is taken to 
be $2^{\circ}$ ($5^{\circ}$). The results for two different values of 
$\theta_{34}$, $2^{\circ}$ and $40^{\circ}$, for $\theta_{13}=8^{\circ}$ and 
vanishing CP phases have been plotted in these figures (the wiggly 
lines). The smooth lines give the predictions of the 
three--neutrino model for $\delta=0$ and  $\theta_{13}^{\,eff}$ as 
given by eq.(\ref{s-13}).  

It is convenient to write in somewhat greater detail 
eq.(\ref{eq:fourfam-e-mu}), 
\bea
P_{3+1} (\nu_e \to \nu_\mu) &=&  4\,s_{14}^2\,s_{24}^2\,c_{34}^4 \sin^2 
\left (\frac {\Delta m^2_{43} L}{4 E} \right ) +
\Big[4 c_{13}^2 s_{13}^2 s_{23}^2 
- \nn \\& & 
      8 \, c_{13}^2 c_{23} s_{13} s_{23} (s_{13}s_{24} + c_{13} s_{23}s_{14}) s_{34}
       \Big]
      \sin^2 \left (\frac {\Delta m^2_{32} L}{4 E} \right )\,+ \label{eq:e-mu} \\& &
      O(s_{14}^2,s_{24}^2,s_{14}s_{24}) \;,\nn 
\eea
where we explicitly show all the involved angles. We also include
the term oscillating with the shortest length, due to the LSND mass difference, 
although it is of higher order in the expansion for small $s_{14}$ and $s_{24}$. 
For small $L$ this term dominates the transition probability and 
it makes the four--neutrino prediction larger than the three--neutrino result, 
where this term does not exist. This effect is numerically relevant for 
large enough gap--crossing angles only, as it may be seen comparing 
fig.~\ref{fig:prob1}a and fig.~\ref{fig:prob1}b.
At intermediate distances ($L \sim $ 3000 Km) the oscillation 
probabilities in the two models are quite similar: at this distance we expect 
therefore that it will be difficult to tell three from four neutrinos. 
At larger distances ( $L > $ 5000 Km) the distinction will in general 
be possible for $\theta_{34}$ large enough.

\begin{figure}[h!]
\begin{center}
\epsfxsize6cm\epsffile{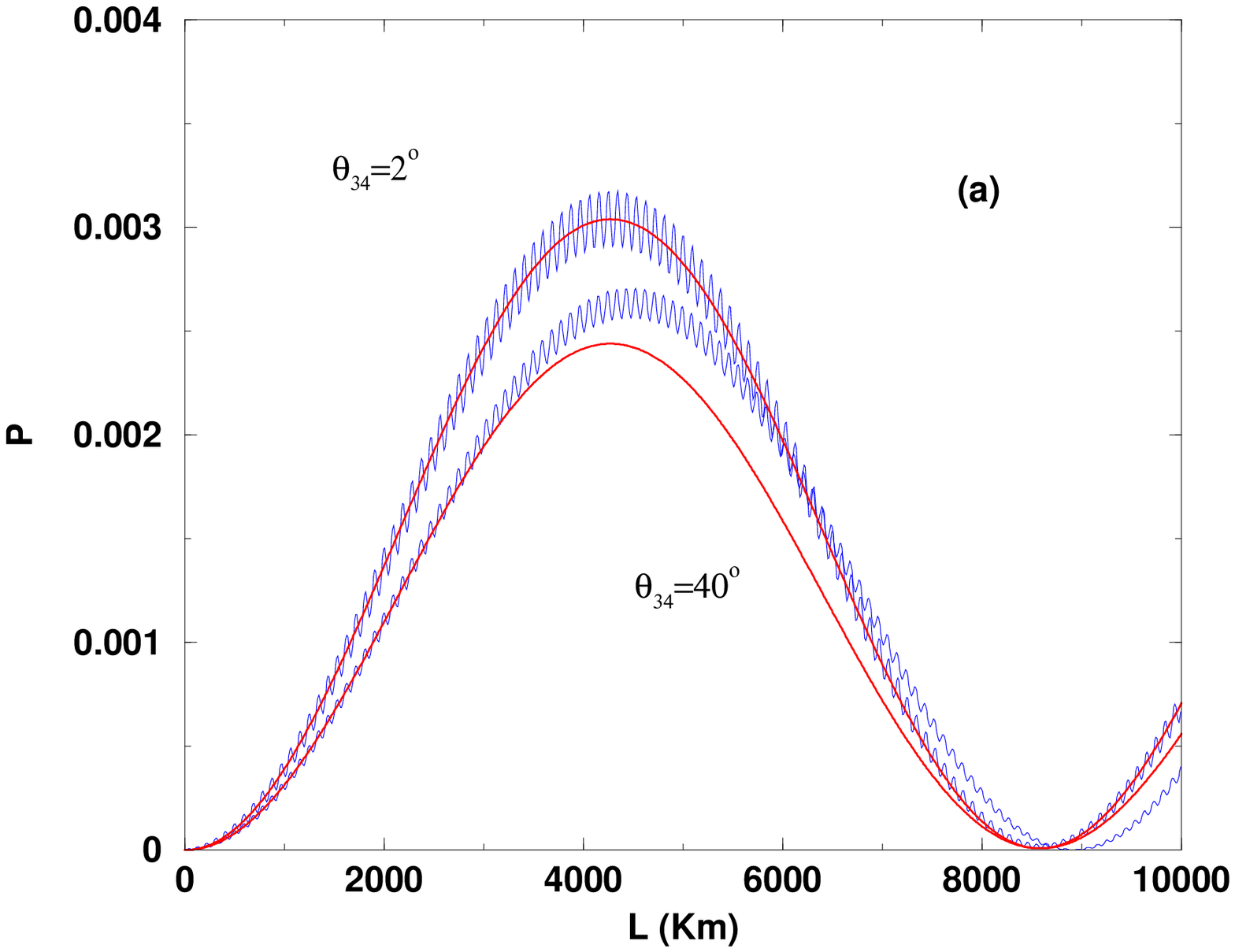} \hspace{0.7truecm} 
\epsfxsize6cm\epsffile{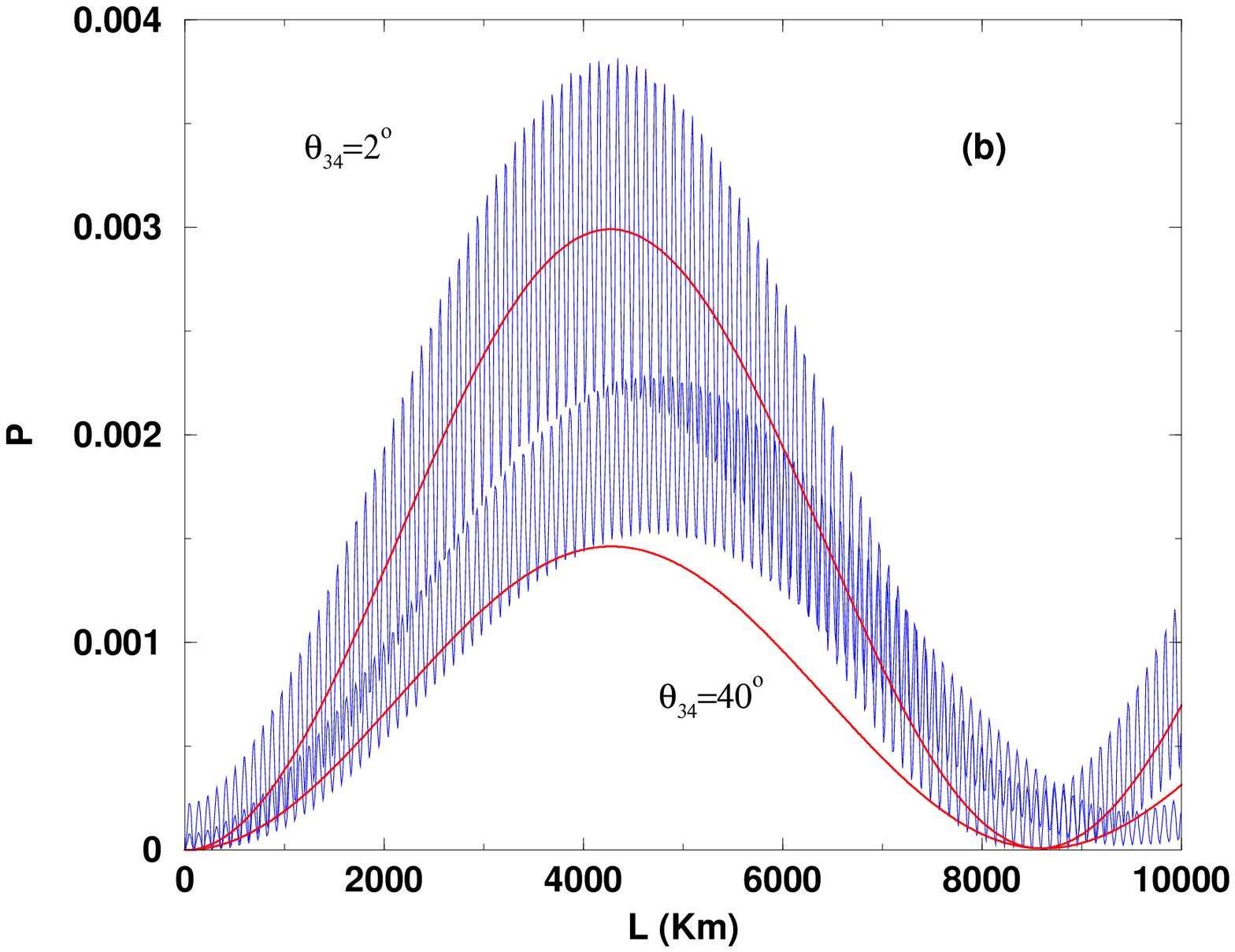}
\caption{ Oscillation probability for $\bar \nu_{e} \to \bar \nu_{\mu}$ as a 
function of the pathlength for $E$ = 38 GeV and (a) $\epsilon 
=2^{\circ}$ or (b) $\epsilon =5^{\circ}$. See text for further details.
\label{fig:prob2}}
\end{center}
\end{figure} 

In the case of antineutrinos, a similar pattern is observed. The 
term oscillating with the atmospheric mass difference in matter
shows a dip at $L \simeq 8000$ Km, evident in fig.~\ref{fig:prob2}a. 
At this distance the term oscillating with LSND frequency dominates:
the average probability for (3+1)--antineutrinos in matter is non-zero 
for $\epsilon = 5^\circ$ and $\theta_{34} = 40^\circ$, 
see fig.~\ref{fig:prob2}b. We also note that the oscillation 
probability in the antineutrino case is very small and this fact, 
together with the further lowering factor due to the ratio of the 
fluxes and cross sections, makes the total number 
of oscillated events essentially dominated by the neutrino 
contribution.

\begin{figure}[h!]
\begin{center}
\epsfxsize15cm\hspace{2cm}\epsffile{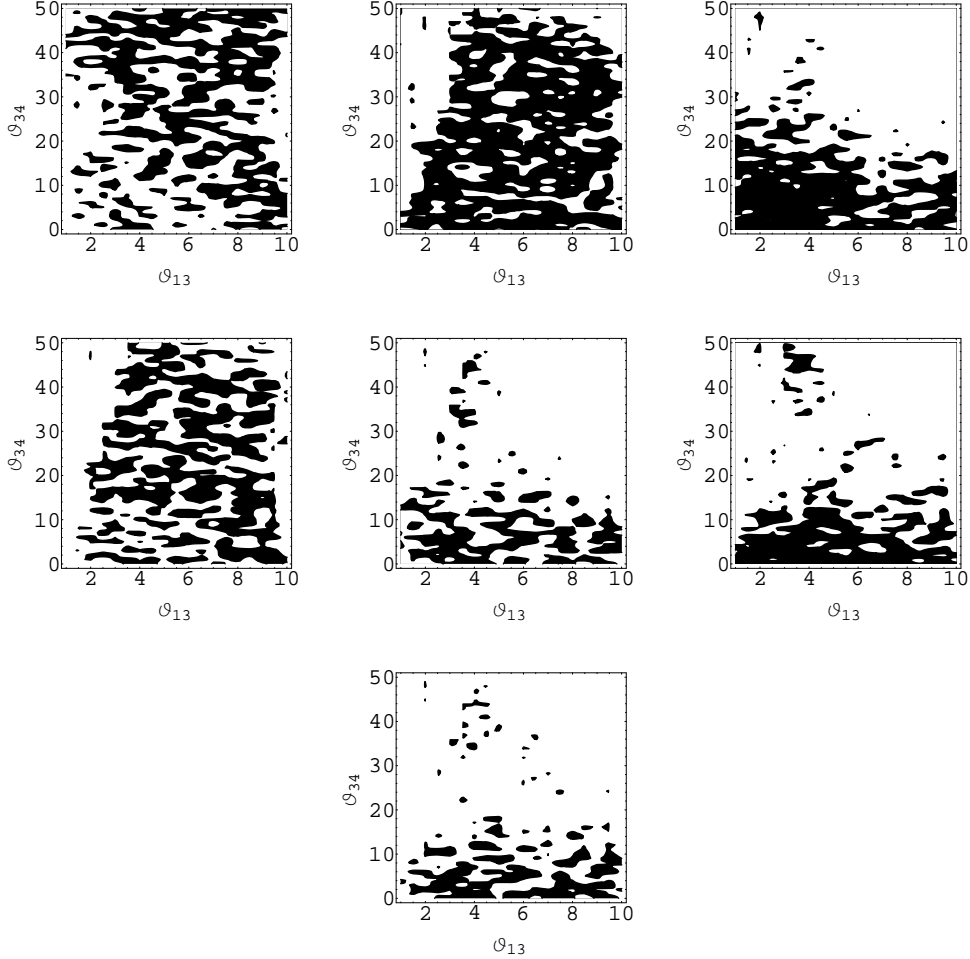} 
\caption{ Plots at 68 \% in the four--family plane for different
 baselines and their combinations for $\theta_{14}=\theta_{24}=2^{\circ}$.
From left to right and top to bottom: 
$\lambda = 1$; $\lambda = 2$; $\lambda = 3$;
$\lambda = 1+2$; $\lambda = 1+3$; $\lambda = 2+3$;
$\lambda = 1+2+3$.
\label{fig:dalm=2}}
\end{center}
\end{figure}

We discuss now the fits that we made to ``data'' generated 
in a (3+1)--neutrino model using a three--neutrino theory. 
The results depend heavily on the values of the small gap-crossing 
angles $\theta_{14}$ and $\theta_{24}$, that we have assumed to be equal 
for simplicity. If their value is small ($2^{\circ}$), since 
the 3+1 model has a smooth limit to the three--neutrino theory,  
the fit is possible for almost every value of the other parameters.
This is shown in fig.~\ref{fig:dalm=2}, where in the dark regions of the 
``dalmatian dog hair'' plot \cite{disney} the three--neutrino model is 
able to fit at 68\%  c.l. the data generated with those  
parameter values. The data include five energy bins, with muons of 
both signs, but the first bin in each polarity has been discarded, 
since the expected efficiency is very low \cite{Cervera:2000vy}.

The reason of the blotted behaviour stands in statistical fluctuations 
in the smearing of the input distributions.
In the first row the data collected at three different distances 
($L$ = 732, 3500 and 7332 Km) are fitted separately: the number of 
degrees of freedom in these cases is 6 (the 68\% c.l. bound is at 
$\chi^{2} = 1.17 \times 6$). Notice that the largest distance gives a 
somewhat better possibility to tell three from four.
In fact, for small values of $\theta_{34}$, the $\nu_e \to \nu_\mu$ 
transition probabilities in matter for both schemes are essentially 
the same whereas for larger $\theta_{34}$ they become different at 
large distances (see fig. \ref{fig:prob1} for $E$ = 38 GeV). 

The second row in fig.~\ref{fig:dalm=2} shows the combined fit to 
two different distances ($\lambda$ = 1 and 2, 1 and 3, 2 and 3, respectively). 
Here the number of degrees of freedom is 14 (the 68\% c.l. bound is at 
$\chi^{2} = 1.14 \times 14$). Notice that the longest baseline causes a 
slight improvement in the discrimination when combined with the other
distances. The last plot refers to using 
data at $\lambda$ = 1, 2 and 3 simultaneously (here we have 22 d.o.f.
and the 68\% c.l. bound at $\chi^{2} = 1.12 \times 22$).

Increasing the number of energy bins from five to ten 
(i.e. assuming a detector with better resolution) for $ \epsilon = 2^\circ$, 
we have not observed a sensible reduction of the blotted regions;
we interpret this result as due to the extremely poor statistics per energy bin
in this case. We do not present the corresponding figure. 

\begin{figure}[h!]
\begin{center}
\epsfxsize13cm\epsffile{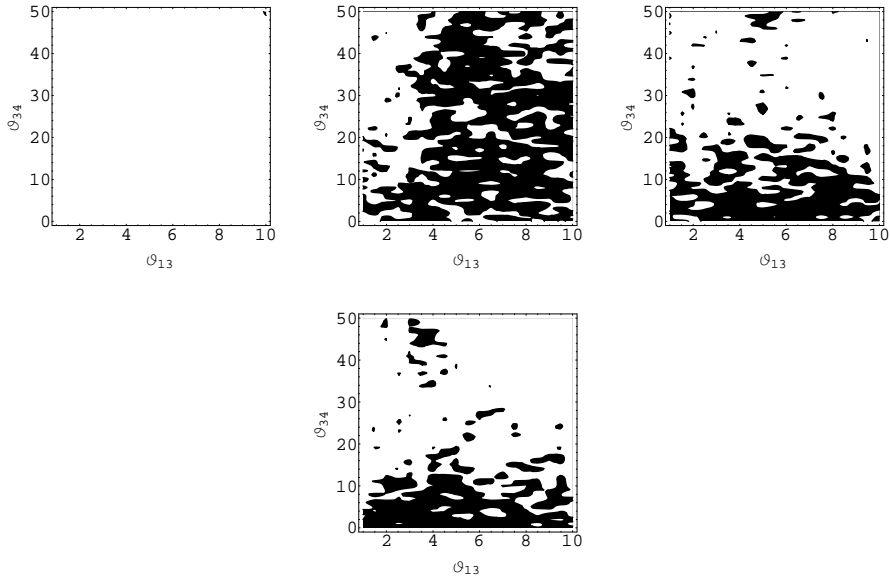} 
\caption{\label{5bin5gradi} Plots at 68 \% in the four--family plane for different
 baselines for $\theta_{14}=\theta_{24}=5^\circ$.
From left to right and top to bottom: 
$\lambda = 1$; $\lambda = 2$; $\lambda = 3$; $\lambda = 2+3$.
\label{fig:dalm=5}}
\end{center}
\end{figure}

On the contrary, increasing the value of $\theta_{14}$ and $\theta_{24}$, 
the extension of blotted regions decreases. 
This is shown for $\theta_{14}=\theta_{24}= 5^{\circ}$ in fig.~\ref{fig:dalm=5}
for five energy bins. In the first row we again present a separate fit for each 
baseline. Notice that for the shortest baseline we can always tell 3 from 3+1. 
Therefore, in the second row we present only the combination of 
$\lambda$ = 2 and 3. 

The interpretation of the results shown in the figure can be easily 
done following our previous considerations and fig.~\ref{fig:prob1}b. At 
the shortest distance, the term in the oscillation probability 
varying with the atmospheric mass difference becomes negligible, and 
the LSND term dominates: the 3+1 model gives a probability larger than 
the three--neutrino theory and confusion is not possible. 
The oscillation probabilities at the 
intermediate distance are very similar in the two models, and the 
confusion is therefore maximal in this case.

\begin{figure}[h!]
\begin{center}
\epsfxsize13cm\epsffile{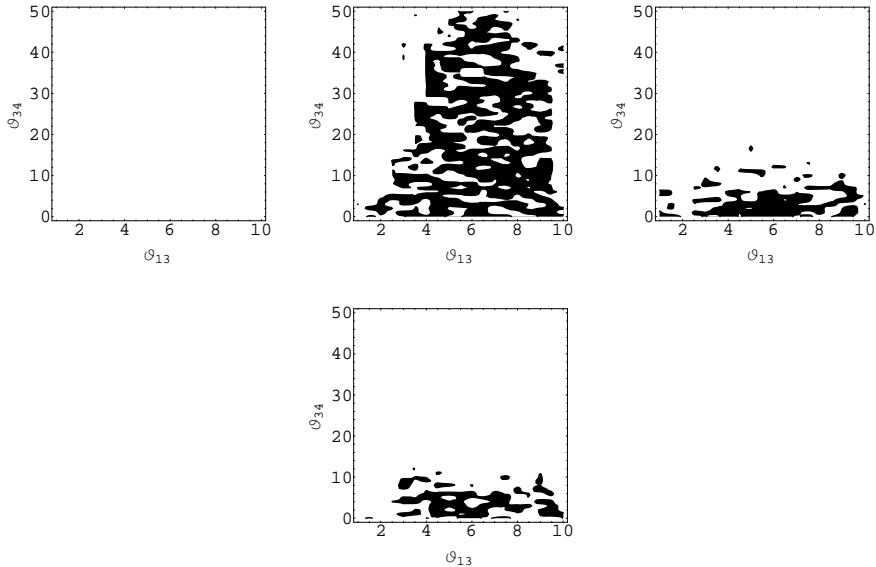} 
\caption{\label{10bin5gradi} Plots at 68 \% in the four--family plane for 
different baselines for ten bins and $\theta_{14}=\theta_{24}=5^\circ$.
From left to right and top to bottom: 
$\lambda = 1$; $\lambda = 2$; $\lambda = 3$; $\lambda = 2+3$.
\label{fig:dalm=5bin10}}
\end{center}
\end{figure}

This situation might be improved if a larger number of energy bins 
could be attained, in that the different energy 
dependence in matter would help in the distinction of the two models.
In fig.~\ref{fig:dalm=5bin10} we present the ``dalmatian dog hair'' plot
when the ``data'' consist of ten energy bins. The regions 
of confusion in this case are considerably reduced. 
As a consequence, a fit with the wrong theory to the largest baseline 
data is only possible for low ($\aprle 10^{\circ}$) values of $\theta_{34}$.

A further increase of the gap--crossing angles to $\theta_{14} = 
\theta_{24} = 10^{\circ}$ makes the distinction (even at 95\% c.l.) of the 
two models possible for all values of the other, variable parameters.

\begin{table} [h!]
{\small
\center 
\begin{tabular}{||c|c|c|c|c|c|c|c||}
\hline\hline 
         & $\lambda = 1$ &$\lambda = 2$ &$\lambda = 3$ &$\lambda = 1+2$ 
         & $\lambda = 1+3$ &$\lambda = 2+3$ &$\lambda = 1+2+3$ \\
\hline\hline
 5 bins               & & & & & & & \\
 $\epsilon = 2^\circ$ & 44 \% & 54 \% & 33 \% & 42 \% & 22 \% & 27 \% & 20 \% \\
 fig. \ref{fig:dalm=2} & & & & & & & \\
\hline
 5 bins               & & & & & & & \\
 $\epsilon = 5^\circ$ & - & 54 \% & 32 \% & - & - & 26 \% & - \\
 fig. \ref{fig:dalm=5} & & & & & & & \\
\hline
 10 bins               & & & & & & & \\
 $\epsilon = 5^\circ$ & - & 36 \% & 11 \% & - & - & 8 \% & - \\
 fig. \ref{fig:dalm=5bin10} & & & & & & & \\
\hline\hline
\end{tabular}
\caption{ Fraction of ``data'' points generated with a four--family
theory that can be fitted with a three--neutrino model.}
\label{tab:percentage}
}
\end{table}

A quantitative feeling of the possibility of confusion (or, with a 
more optimistic attitude, of discrimination) can be given by the 
numbers reported in Table 1. They correspond to the fraction of 
points, among the 969 ``data'' sets fitted, in which the three-neutrino 
model can give a good fit (at 68\% c.l.). The same points  
have been smoothly joined to obtain the ``dalmatian dog hair'' plots in 
figs.~\ref{fig:dalm=2}-\ref{fig:dalm=5bin10}.  The empty cells in the 
table correspond to situations of no possible confusion.

\begin{figure}[h!]
\begin{center}
\epsfxsize15cm\epsffile{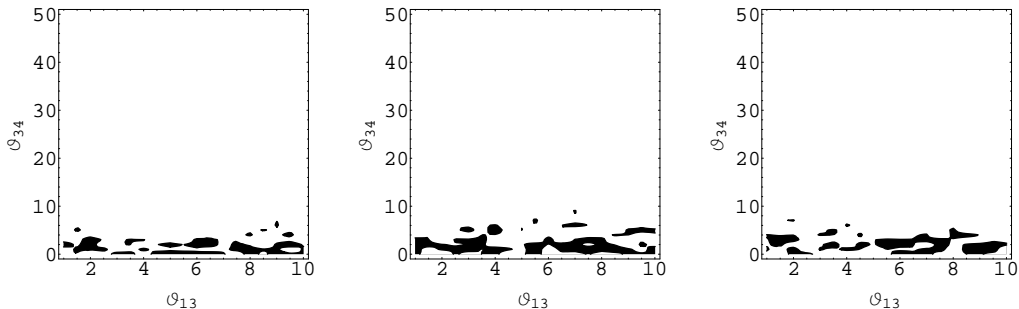} 
\caption{Plots at 68\% c.l. in the four--family plane for different
 baselines in the $\nu_\mu \to \nu_\tau$ channel, for 
 $\theta_{14}=\theta_{24}=2^\circ$.
From left to right: 
$\lambda = 1$; $\lambda = 2$; $\lambda = 3$.
\label{fig:mutau68}}
\end{center}
\end{figure}

To complete our discussion regarding the 3+1 scheme against 
the three--family theory we consider now the $\nu_\mu \to \nu_\tau$ transition. 
We notice that the three--family probability is not recovered
for $\theta_{14} = \theta_{24} = \epsilon \to 0$. 
In this limit the four--family transition probability  
equals the three--family one times a correction factor $c_{34}^2$, see 
eq.(\ref{eq:fourfam-mu-t}): 
\be
P_{3+1} (\nu_\mu \to \nu_\tau) \to c_{34}^2 \, P_3 (\nu_\mu \to \nu_\tau)\;.
\ee 
Due to unitarity, we also have
\be
P_{3+1} (\nu_\mu \to \nu_s) \to s_{34}^2 \, P_3 (\nu_\mu \to \nu_\tau)\;.
\ee 
The region where confusion between the 3+1 model and the three--family theory
is possible is that of low $\theta_{34}$. 
In this case, we considered  
an idealized detector with constant efficiency (35\%) and a very low level 
background (10$^{-5}$). The mass of the detector is 4 Kton and again 
five energy bins have been assumed. 
In fig.~\ref{fig:mutau68} we show the 68$\%$ c.l. plots for this channel, 
where the expected behaviour is clearly observable, for any of the considered
baselines.
 
Up to this point, we have only discussed in this Section the 3+1 
scheme. We noticed in the previous Section that it is much more difficult  
to reproduce ``data'' generated in the 2+2 four neutrino scheme with 
the three-family theory. 
We tried to do this in much the same manner as in the 3+1 
case. The result can be described 
giving the numbers corresponding to the entries of the previous Table, 
that are all smaller than 5\%. This maximum value is obtained 
for the intermediate distance, $L = 3500$ Km, and corresponds to a 
narrow region for quite small values of all the gap--crossing angles.
Therefore we do not present the relative ``dalmatian'' plots, which in 
this case are affected by albinism. 

\section{CP violation vs. more neutrinos}
\label{sec:delta}

In the previous Section, we have extensively illustrated when confusion
between the three--neutrino and the (3+1)--neutrino models is possible. 
However, we have not
presented in detail the results of a fit of the four--family ``data'' 
in the parameter space of the three--family model: 
in fact, the ``dalmatian'' plots in the previous Section represent the 
regions of confusion in four--family parameter space. 

In three families, the free parameters that we considered are one
angle, $\theta_{13}$, and the CP violating phase, $\delta$.
These parameters are expected to be still 
unknown (or poorly known) when the Neutrino Factory will be operative. 
In particular, we are interested here in the following questions: 
\begin{itemize}
    \item is it possible to find a non-vanishing CP violating phase 
    when fitting with the three--family model the 3+1 (CP conserving) ``data''? 
    \item is it possible to fit with a 3+1 (CP conserving) theory the 
    data if their fit in the three--neutrino model gives a large CP 
    violating phase $\delta$ ?
\end{itemize}

\begin{figure}[h!]
\begin{center}
\begin{tabular}{ccc}
\hspace{-1cm} \epsfxsize5.5cm\epsffile{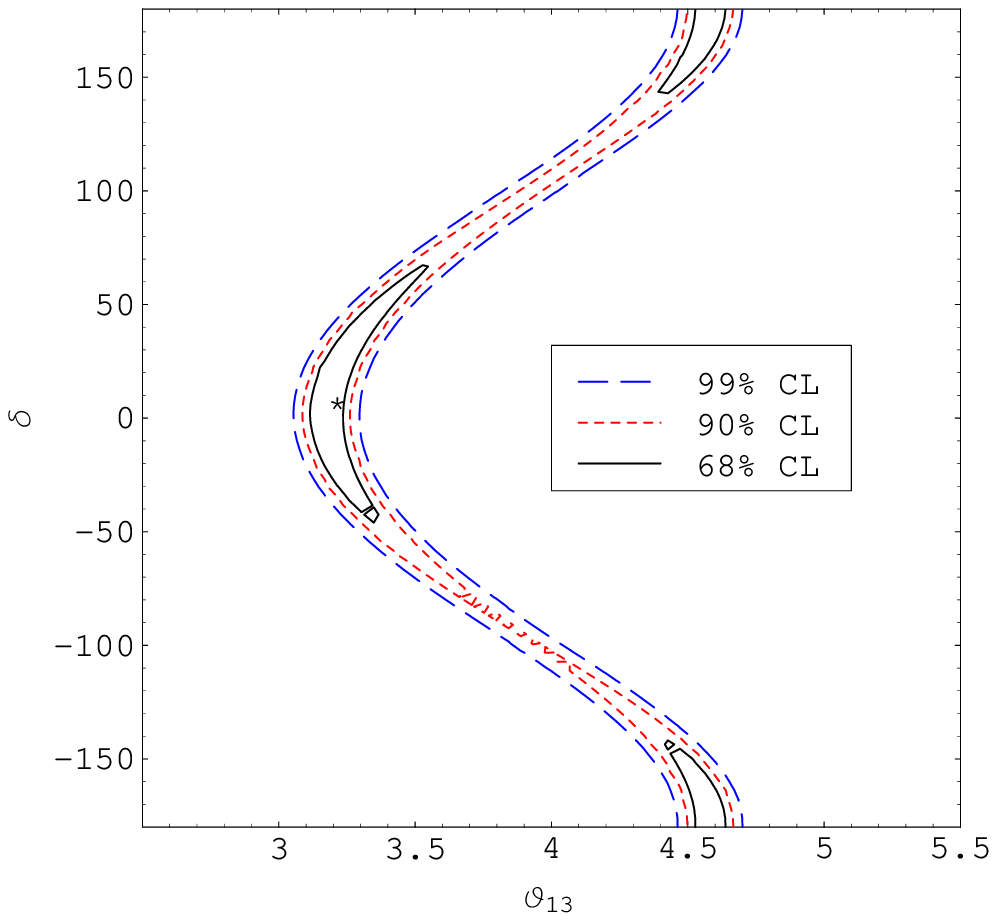} &
\epsfxsize5.5cm\epsffile{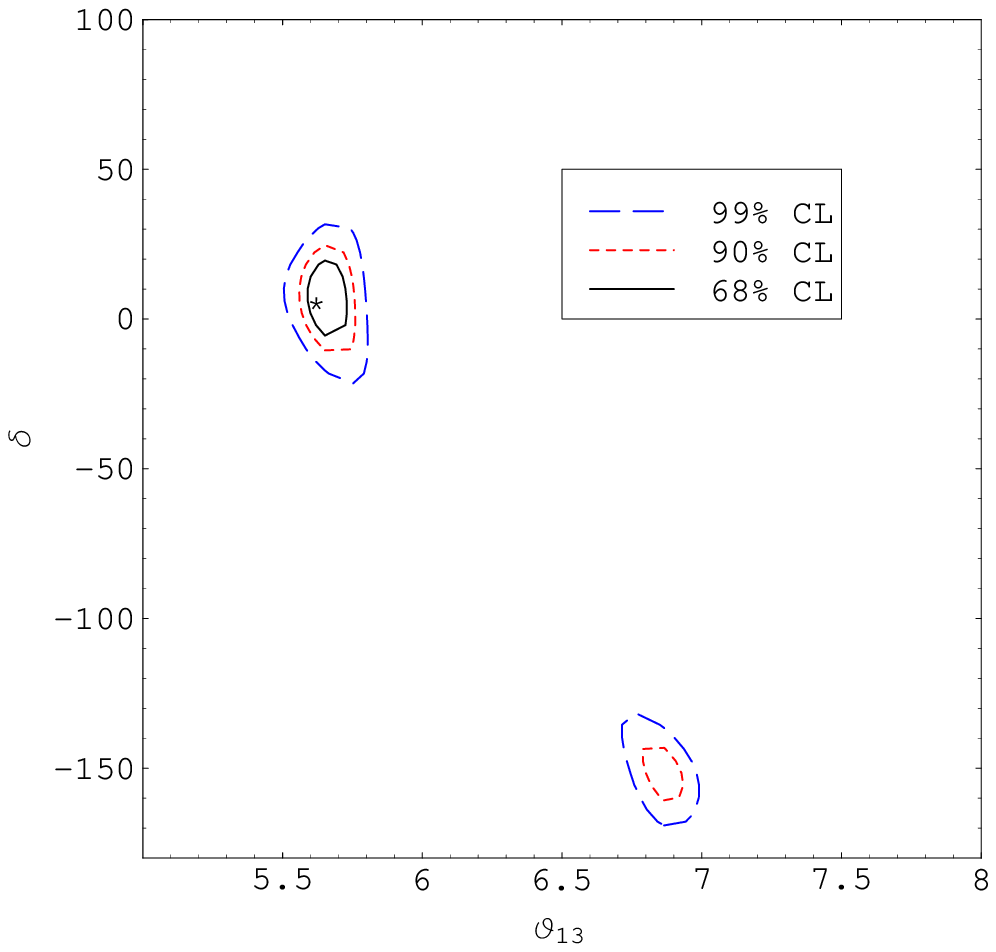} &
\epsfxsize5.5cm\epsffile{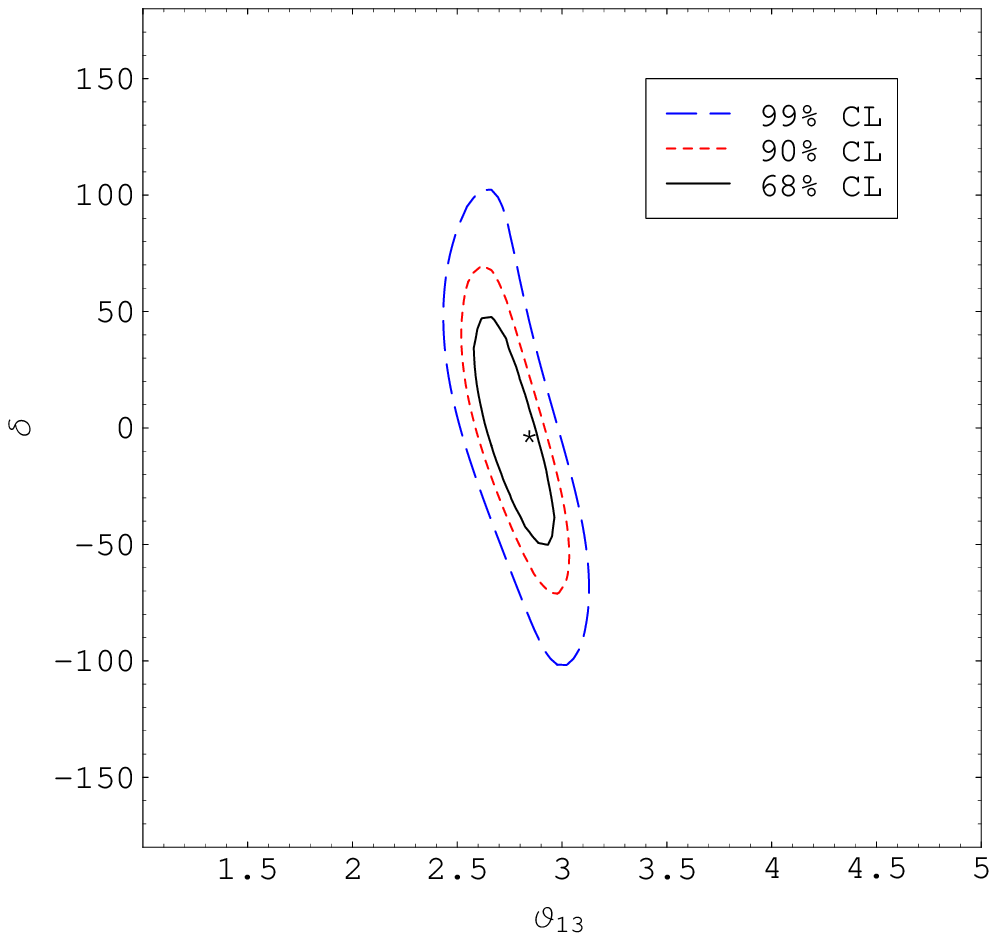}
\end{tabular}
\caption{\label{fig:3-CP-fit}
Confidence level contours for typical points where the 
three--neutrino theory can well reproduce (3+1)--neutrino ``data'' at the 
three distances studied: from left to right, $L=732$, 3500 and 7332 Km.}
\end{center}
\end{figure}

In order to answer the first question, we present in fig.~\ref{fig:3-CP-fit}
the 68\%, 90\% and 99\% confidence level contours 
\footnote{The c.l. contours correspond to 
$\Delta\chi^2=2.30,\;4.61,\;9.21$, 
respectively, since there are two free parameters. 
The best fit points are represented by stars.}
for typical three--neutrino fits (one for each distance) 
to 3+1 neutrino ``data'' 
generated with $\theta_{14}=\theta_{24}=2^\circ$ and five energy bins. 
The parameters are best determined for the intermediate distance, 3500 
Km, that is close to the optimal distance for studies of CP violation in 
the three--neutrino scenario
\cite{Cervera:2000kp,Barger:2000yn,Bueno:2000fg,Freund:2001ui}. 
In fact, the value of the fitted phase $\delta$ stays close to zero, 
except for a possible `mirror' region around $- 160^\circ$. The 
existence of such mirror regions has been discussed in  
\cite{Burguet-Castell:2001ez,Freund:2001ui}: they may be removed 
increasing the energy resolution \cite{Rubbia:2001pk} and/or combining data 
at two different distances. At the largest distance studied, 7332 Km, 
the uncertainty in the value of $\delta$ is higher, since the matter 
effects dominate over the possible effect of CP 
violation in the oscillations and  $\delta$ is less easily determined.  
Finally, at the shortest distance and at 99\% c.l. any value of $\delta$ 
is allowed. Plots at this distance with a similar behaviour 
have been also presented in \cite{Burguet-Castell:2001ez}: 
this distance is clearly not suited for a precise determination 
of the CP violating phase, at least for the set-up of 
Neutrino Factory and detector considered here.
To illustrate the representativeness of the values chosen 
for the plots in fig.~\ref{fig:3-CP-fit} we note 
that, out of about 6000 successful fits, 31\% (50\%, 37\%) for 
$\lambda =$ 1 (2, 3) give for the CP violating phase a value 
$-15^\circ < \delta < 15^\circ$. 

\begin{figure}[ht!]
\begin{center}
\hspace{-1cm} 
\epsfxsize15cm\epsffile{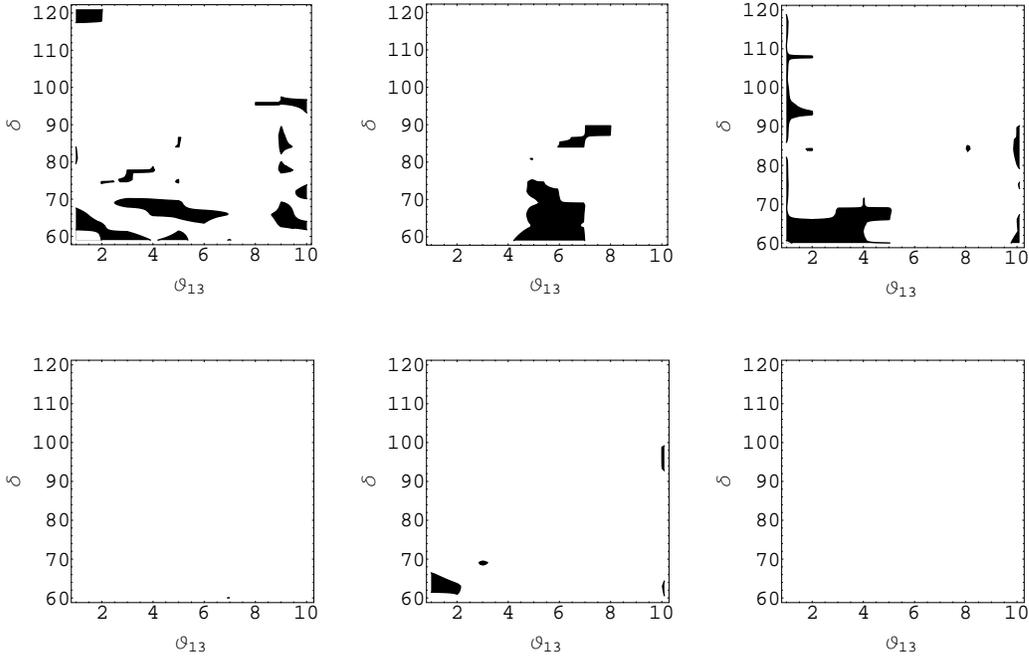}
\caption{Examples of regions in the ($\theta_{13}$, $\delta$) plane  
where the (3+1)--neutrino theory with vanishing CP violating phases can 
reproduce at 90\% c.l. three--neutrino ``data''.
From left to right and top to bottom: 
$\lambda = 1$; $\lambda = 2$; $\lambda = 3$; $\lambda = 1+2$;
$\lambda = 1+3$; $\lambda = 2+3$. \label{fig:3in4}}
\end{center}
\end{figure}

In order to answer the second question, and as a cross--check of our 
previous results, we adopted the converse procedure, 
namely we fitted ``data'' 
generated in a three-neutrino, CP violating, scenario with the formulae 
of the 3+1 model with CP violating phases fixed to zero. We chose for 
this purpose to vary the three--neutrino 
parameters in a region where the confusion is often not possible 
in the previous fit, $1^\circ \leq \theta_{13} \leq 10^\circ$ and
$60^\circ \leq \delta \leq 120^\circ$. 
This region has only a 3\% ($<$0.02\%, 1.5\%) of successful fits falling in it
for $\lambda$ = 1 (2,3). 
Among the parameters in the 3+1 
theory, $\theta_{13}$ and $\theta_{34}$ have been left free to vary, while 
$\theta_{14}=\theta_{24}=\epsilon$ have been kept fixed to
$2^\circ, 5^\circ$ and 10$^\circ$. 
The corresponding plots for five energy bins and for  
$\epsilon = 2^\circ$ are 
presented in fig.~\ref{fig:3in4}, where the dark regions 
correspond to an acceptable fit (at 90\% c.l.) in the 3+1 theory. 

The results are precisely what was expected on the basis of the 
previous fits. The fit 
with the wrong model, with the chosen values of parameters, is 
possible in many a case for $L$ = 732 Km, it is very difficult to 
obtain for the intermediate distance, $L$ = 3500 Km, and at the largest $L$
an intermediate situation holds. Fitting simultaneously data at two different 
distances the possibility of confusion is strongly reduced and in fact 
it vanishes if the data at intermediate distance are used in any 
combination with the others. We also observed that no confusion would be 
possible assuming the larger values $\epsilon$ = 5$^\circ$ or 
10$^\circ$. 

Our result seems to imply that if the data would point to a maximal 
CP violating phase in the three--neutrino theory, it would be very 
difficult to describe them in a theory without CP violation, 
even if with more neutrinos.  

\section{Conclusions}
\label{sec:concl}

The present experimental data for solar and atmospheric neutrinos
give quite strong indications in favour of neutrino oscillations and 
of nonzero neutrino masses. In addition, the results of the LSND 
experiment would imply the existence
of a puzzling fourth, sterile, neutrino state. 
The comprehensive analysis of all the other data (including the recent 
SNO results) seems however to disfavour both possible classes of 
four--neutrino spectra, the so-called 2+2 and 3+1 schemes. 

An experimental set-up with the ambitious goal of precision
measurement of the whole three-neutrino mixing parameter space
(including the extremely important quest for leptonic CP violation)
is under study. This experimental programme consists of the development
of a ``Neutrino Factory'' (high-energy muons decaying in the straight section 
of a storage ring and producing a very pure and intense neutrino beam) 
and of suitably optimized detectors located very far away from the 
neutrino source. 
The effort to prepare such very long baseline neutrino experiments
will require a time period covering this and the beginning of the following 
decade, in the absence of a conclusive confirmation of the LSND results (that 
would call instead for a short baseline experiment to investigate 
the four-neutrino mixing parameter space). 

It is of interest to explore the capability of such an 
experimental set-up to discriminate between a three-neutrino
model and a possible scenario where a fourth neutrino is included. 
Some relevant questions came to our mind. 
Can the experimental data be described equally well with a three--neutrino 
theory or a four--neutrino model? Even worse, is it possible
for a CP-conserving three active and one sterile neutrino model
to be confused with a CP-violating three (active) neutrino theory?

In this paper we tried to answer these questions 
in the framework of an LSND inspired four neutrino model, assuming 
a ``large'' squared mass difference $\Delta m^2 \simeq 1\;{\rm eV}^2$. 
However, the approach is in principle independent of the choice
of the third mass gap (in addition to the solar and
atmospheric ones) and can be easily repeated for different
values of $\Delta m^2$. 

The results of our analysis can be summarized as follows. 
Data that could be described in four--family without CP violation can 
also be fitted with the three--family formulae in some particular 
zones of the four--family parameter space, not restricted 
to the obvious case of very small angles. These zones are reduced 
in size for increasing gap--crossing angles and energy resolution of 
the detector. The use of $\nu_{\mu} \to \nu_\tau$ oscillations 
through direct or indirect detection of tau lepton decays would allow 
a much easier discrimination of the two theoretical hypotheses. 

In the zones of the four--family parameter space 
that can be also described with the three-family 
theory, the fitted value of the CP violating phase, $\delta$, 
is generally not large. In particular, this is true for $L = 3500$ Km, 
whereas for $L = 7332$ Km the determination of $\delta$ is somewhat looser. 
For the shortest baseline, $L = 732$ Km, we have the largest spread in the 
values of $\delta$, although the most probable value is still close to zero. 

Data that can be fitted with a CP phase close to 90$^\circ$ in the 
three--neutrino theory cannot be described  
in a CP conserving 3+1 theory, provided that   
data at two different distances are used.  

Finally, in the 2+2 scheme (as opposed to 3+1) the ambiguity with a 
three--neutrino theory is essentially absent. 

\vspace{-0.5cm}
\section*{Acknowledgements}

We acknowledge useful conversations with M.B. Gavela, 
P. Hernandez, C. Pe\~na-Garay and J. Sato.
We are particularly indebted with P. Lipari for discussions 
on many different aspects of this paper.


\newpage

\begin{thebibliography}{999}

\bibitem{Pontecorvo:1957yb}
B.~Pontecorvo,
Sov.\ Phys.\ JETP {\bf 6} (1957) 429
[Zh.\ Eksp.\ Teor.\ Fiz.\  {\bf 33} (1957) 549].

\bibitem{Maki:1962mu}
Z.~Maki, M.~Nakagawa and S.~Sakata,
Prog.\ Theor.\ Phys.\ {\bf 28} (1962) 870.

\bibitem{Pontecorvo:1968fh}
B.~Pontecorvo,
Sov.\ Phys.\ JETP {\bf 26} (1968) 984.

\bibitem{Gribov:1969kq}
V.~N.~Gribov and B.~Pontecorvo,
Phys.\ Lett.\ B {\bf 28} (1969) 493.

\bibitem{Cleveland:1998nv}
B.~T.~Cleveland {\it et al.},
Astrophys.\ J.\  {\bf 496} (1998) 505.

\bibitem{Fukuda:1996sz}
Y.~Fukuda {\it et al.}  [Kamiokande Collaboration],
Phys.\ Rev.\ Lett.\  {\bf 77} (1996) 1683.

\bibitem{Hampel:1999xg}
W.~Hampel {\it et al.}  [GALLEX Collaboration],
Phys.\ Lett.\  {\bf B447} (1999) 127.

\bibitem{Abdurashitov:1999zd}
J.~N.~Abdurashitov {\it et al.}  [SAGE Collaboration],
Phys.\ Rev.\  {\bf C60} (1999) 055801.

\bibitem{Suzuki:1999cy}
Y.~Suzuki  [Super-Kamiokande Collaboration],
Nucl.\ Phys.\ Proc.\ Suppl.\  {\bf 77} (1999) 35.

\bibitem{Ahmad:2001an}
Q.~R.~Ahmad {\it et al.}  [SNO Collaboration],
nucl-ex/0106015.

\bibitem{Fukuda:1994mc}
Y.~Fukuda {\it et al.}  [Kamiokande Collaboration],
Phys.\ Lett.\  {\bf B335} (1994) 237.

\bibitem{Becker-Szendy:1995vr}
R.~Becker-Szendy {\it et al.},
Nucl.\ Phys.\ Proc.\ Suppl.\  {\bf 38} (1995) 331.

\bibitem{Fukuda:1999ah}
Y.~Fukuda {\it et al.}  [SuperKamiokande Collaboration],
Phys.\ Rev.\ Lett.\  {\bf 82} (1999) 2644.

\bibitem{Allison:1999ms}
W.~W.~Allison {\it et al.}  [Soudan-2 Collaboration],
Phys.\ Lett.\  {\bf B449} (1999) 137.

\bibitem{Ambrosio:1998wu}
M.~Ambrosio {\it et al.}  [MACRO Collaboration],
Phys.\ Lett.\  {\bf B434} (1998) 451.

\bibitem{Toshito:2001dk}
T.~Toshito  [SuperKamiokande Collaboration],
hep-ex/0105023.

\bibitem{Athanassopoulos:1998pv}
C.~Athanassopoulos {\it et al.}  [LSND Collaboration],
Phys.\ Rev.\ Lett.\  {\bf 81} (1998) 1774.

\bibitem{Aguilar:2001ty}
A.~Aguilar  [LSND Collaboration],
hep-ex/0104049.

\bibitem{Kleinfeller:2000em}
J.~Kleinfeller  [KARMEN Collaboration],
Nucl.\ Phys.\ Proc.\ Suppl.\  {\bf 87} (2000) 281.

\bibitem{Church:1997jc}
E.~Church {\it et al.}  [BooNe Collaboration],
nucl-ex/9706011.

\bibitem{Caso:2000tc}
C.~Caso and A.~Gurtu,
Eur.\ Phys.\ J.\ C {\bf 15} (2000) 256.

\bibitem{Fogli:1999zq}
G.~L.~Fogli, E.~Lisi, A.~Marrone and G.~Scioscia,
hep-ph/9906450.

\bibitem{Geer:1998iz}
S.~Geer,
Phys.\ Rev.\ D {\bf 57} (1998) 6989
[Erratum-ibid.\ D {\bf 59} (1998) 039903]
[hep-ph/9712290].

\bibitem{DeRujula:1999hd}
A.~De Rujula, M.~B.~Gavela and P.~Hernandez,
Nucl.\ Phys.\  {\bf B547} (1999) 21.

\bibitem{Barger:2000fs}
V.~Barger, S.~Geer and K.~Whisnant,
Phys.\ Rev.\ D {\bf 61} (2000) 053004
[hep-ph/9906487].

\bibitem{Bueno:2000wb}
A.~Bueno, M.~Campanelli and A.~Rubbia,
Nucl.\ Phys.\ B {\bf 573} (2000) 27.

\bibitem{Dick:1999ed}
K.~Dick, M.~Freund, M.~Lindner and A.~Romanino,
Nucl.\ Phys.\ B {\bf 562} (1999) 29
[hep-ph/9903308].

\bibitem{Cervera:2000kp}
A.~Cervera {\it et al.},
Nucl.\ Phys.\  {\bf B579} (2000) 17.

\bibitem{Albright:2000xi}
C.~Albright {\it et al.},
hep-ex/0008064.

\bibitem{Blondel:2000gj}
A.~Blondel {\it et al.},
Nucl.\ Instrum.\ Meth.\ A {\bf 451} (2000) 102.

\bibitem{Barger:2001qd}
V.~Barger {\it et al.},
hep-ph/0103052.

\bibitem{Donini:1999jc}
A.~Donini, M.~B.~Gavela, P.~Hernandez and S.~Rigolin,
Nucl. Phys. {\bf 574} (2000) 23; 
Nucl.\ Instrum.\ Meth.\ A {\bf 451} (2000) 58
[hep-ph/9910516].

\bibitem{Donini:2001xy}
A.~Donini and D.~Meloni,
hep-ph/0105089.

\bibitem{Bilenkii:1999ny}
S.~M.~Bilenkii, C.~Giunti, W.~Grimus and T.~Schwetz,
Phys.\ Rev.\  {\bf D60} (1999) 073007.

\bibitem{Dydak:1984zq}
F.~Dydak {\it et al.},
Phys.\ Lett.\ B {\bf 134} (1984) 281.

\bibitem{Stockdale:1985ce}
I.~E.~Stockdale {\it et al.},
Z.\ Phys.\ C {\bf 27} (1985) 53.

\bibitem{Declais:1995su}
Y.~Declais {\it et al.},
Nucl.\ Phys.\  {\bf B434} (1995) 503.

\bibitem{Eskut:2001de}
E.~Eskut {\it et al.}  [CHORUS Collaboration],
Phys.\ Lett.\ B {\bf 497} (2001) 8.

\bibitem{Astier:2001yj}
P.~Astier {\it et al.}  [NOMAD Collaboration],
decays,''
hep-ex/0106102.

\bibitem{Barger:2000ch}
V.~Barger, B.~Kayser, J.~Learned, T.~Weiler and K.~Whisnant,
Phys.\ Lett.\ B {\bf 489} (2000) 345
[hep-ph/0008019].

\bibitem{Yasuda:2000xs}
O.~Yasuda,
hep-ph/0102166.

\bibitem{Giunti:2001ur}
C.~Giunti and M.~Laveder,
JHEP{\bf 0102} (2001) 001
[hep-ph/0010009].

\bibitem{Peres:2001ic}
O.~L.~Peres and A.~Y.~Smirnov,
Nucl.\ Phys.\ B {\bf 599} (2001) 3
[hep-ph/0011054].

\bibitem{Grimus:2001mn}
W.~Grimus and T.~Schwetz,
hep-ph/0102252.

\bibitem{Gonzalez-Garcia:2001hs}
M.~C.~Gonzalez-Garcia and C.~Pena-Garay,
Phys.\ Rev.\ D {\bf 63} (2001) 073013
[hep-ph/0011245].

\bibitem{Gonzalez-Garcia:2001uy}
M.~C.~Gonzalez-Garcia, M.~Maltoni and C.~Pena-Garay,
hep-ph/0105269.

\bibitem{Barger:2001zs}
V.~Barger, D.~Marfatia and K.~Whisnant,
hep-ph/0106207.

\bibitem{Bahcall:2001zu}
J.~N.~Bahcall, M.~C.~Gonzalez-Garcia and C.~Pena-Garay,
hep-ph/0106258.

\bibitem{DeRujula:1980yy}
A.~De Rujula, M.~Lusignoli, L.~Maiani, S.~T.~Petcov and R.~Petronzio,
Nucl.\ Phys.\ B {\bf 168} (1980) 54.

\bibitem{Cervera:2000vy}
A.~Cervera, F.~Dydak and J.~Gomez Cadenas,
Nucl.\ Instrum.\ Meth.\ A {\bf 451} (2000) 123.

\bibitem{Donini:2000ky}
A.~Cervera {\it et al.},
hep-ph/0007281.

\bibitem{Burguet-Castell:2001ez}
J.~Burguet-Castell, M.~B.~Gavela, J.~J.~Gomez-Cadenas, P.~Hernandez and O.~Mena,
hep-ph/0103258.

\bibitem{Fogli:2001ir}
G.~L.~Fogli, E.~Lisi and A.~Marrone,
Phys.\ Rev.\ D {\bf 63} (2001) 053008
[hep-ph/0009299].

\bibitem{PREM}
A.~M.~Dziewonski and D.~L.~Anderson, Phys. Earth Planet. Inter. {\bf 25}
(1981) 297.

\bibitem{Ota:2001hf}
T.~Ota and J.~Sato,
Phys.\ Rev.\ D {\bf 63} (2001) 093004
[hep-ph/0011234].

\bibitem{disney}
W. Disney {\sl et al.}, ``One hundred and one dalmatians'', 1961.

\bibitem{Barger:2000yn}
V.~Barger, S.~Geer, R.~Raja and K.~Whisnant,
Phys.\ Rev.\ D {\bf 62} (2000) 073002
[hep-ph/0003184].

\bibitem{Bueno:2000fg}
A.~Bueno, M.~Campanelli and A.~Rubbia,
Nucl.\ Phys.\ B {\bf 589} (2000) 577
[hep-ph/0005007].

\bibitem{Freund:2001ui}
M.~Freund, P.~Huber and M.~Lindner,
hep-ph/0105071.

\bibitem{Rubbia:2001pk}
A.~Rubbia,
hep-ph/0106088.

\end{thebibliography}
\end{document}